\documentclass{jpsj2}
\usepackage{color}
\title{
Exact Scale Invariance in Mixing of Binary Candidates in Voting Model
}

\author{Shintaro \textsc{Mori}$^{1}$\thanks{E-mail :
mori@sci.kitasato-u.ac.jp} and Masato
\textsc{Hisakado}$^{2}$\thanks{E-mail: 
masato\_hisakado@standardandpoors.com}}

\inst{$^{1}$
Department of Physics, School of Science, Kitasato University
\\ Kitasato 1-15-1, Sagamihara, Kanagawa 228-8555 \\
$^{2}$
Standard and Poor's, Marunouchi 1-6-5, Chiyoda-ku, Tokyo 100-0005
}

\abst{
We introduce a voting model and discuss the 
scale invariance in the mixing 
of candidates.  
The Candidates are classified into two
 categories $\mu\in \{0,1\}$ and  are called as `binary' candidates.
There are in total
$N=N_{0}+N_{1}$ candidates, 
and voters vote for them one by one.
The probability that a candidate gets a vote is proportional 
 to the number of votes. The initial number of votes (`seed')
of a candidate $\mu$ is set to be $s_{\mu}$.
After infinite counts of voting, the probability function of the
 share of 
 votes of the candidate $\mu$
obeys gamma distributions with the shape exponent $s_{\mu}$
in the thermodynamic limit
 $Z_{0}=N_{1}s_{1}+N_{0}s_{0}\to \infty$.
Between the cumulative functions $\{x_{\mu}\}$ 
of binary candidates, the  power-law relation 
$1-x_{1} \sim  (1-x_{0})^{\alpha}$ 
with the critical exponent $\alpha=s_{1}/s_{0}$ 
holds
in the region $1-x_{0},1-x_{1}<<1$. 
In the double scaling limit $(s_{1},s_{0})\to (0,0)$ and 
$Z_{0} \to \infty$ with $s_{1}/s_{0}=\alpha$ fixed, 
the relation $1-x_{1}=(1-x_{0})^{\alpha}$ holds
 exactly over the entire range $0\le x_{0},x_{1} \le 1$. 
We study the data on horse races obtained  from
the Japan Racing Association for the period 
 1986 to 2006 and confirm scale invariance.
}

\kword{
scale invariance, voting model, branching process,
gamma distribution, ROC, accuracy ratio
}

\begin{document}
\maketitle

\section{Introduction} 
Scale-invariant behaviour has attracted considerable attention on 
account of its ubiquity in natural and man-made phenomena \cite{Newman}.
  Many possible candidate mechanisms that gives rise to power-law
  distributions have been proposed thus far.
 The Yule process is a widely applicable mechanism 
for generating power-law distributions \cite{Yule}. 
Originally, it has been proposed to
explain why the distribution of the number of species in a genus, a
family, 
or any other taxonomic group follows a power law \cite{Willis}.
 Now, it has found wide applications in other areas \cite{Newman,
Barabasi}.

Consider the distribution of the number of species in a genus.
Suppose first that new species appear but they never die; species are
only ever added to genera and never removed. Species are added to genera
by speciation, the splitting of one species into two. If we assume
that this happens at some stochastically constant rate, then it follows 
 that  a genus with $k$ species
 will gain new species at a rate proportional to $k$,  
since each of the $k$ species has the same chance per unit time of
dividing into two.
In addition, 
suppose that a new species that belongs to  a new genus is added 
once every $m$ speciation events. So $m+1$ new species appear for each
new genus and there are $m+1$ species per genus. Thus the number of
genera goes up steadily as does the number of species within each genus.
We denote the fraction of
genera that has $k$ species by $p_{k,n}$, where $n$ denotes 
the total number of genera and $n$ measures the passage 
of time in the model. At each time-step one new species founds a new
genus, thereby increasing $n$ by 1, and $m$ other species are
 are added to various pre-existing genera which are selected in
 proportion to the number of species they already have.
By solving the  master equation for $p_{k,n}$ in the limit 
$n\to \infty$, $p_{k}\equiv  \lim_{n\to \infty}p_{k,n}$ 
behaves as $p_{k}\sim k^{2+\frac{1}{m}}$.
The Yule process has been adopted and generalized to 
explain power laws in many other systems.
An important feature of this process is that the probability 
that a genus with $k$ species will gain new species is proportional
to $k$. This `rich-get-richer' process is the most important
factor in exhibiting power-law behaviour. The feature that 
 $n$ increases infinitely is also important in generating power-law 
behaviour.

In this study, we introduce a voting model,
 a multivariate Polya-Eggenberger model \cite{Polya,Berg},
and discuss the scale
invariance in the mixing of candidates.
The candidates are classified into two categories $\mu\in \{0,1\}$
and are called as `binary' candidates.
The probability that a candidate get a vote is proportional 
to the number of votes, which is the same as the relation in the  
Yule process.
The main difference between the voting model and the Yule process
is that the number of
candidates is fixed in our model. In the Yule process, $n$ increases and
in  the limit $n\to \infty$, power-law behaviour is observed. 
In our model,   
the distribution of the number of votes   
does not show  power-law behaviour. However, our model exhibits 
 scale-invariant behaviour. This behaviour is observed in the mixing of 
the binary candidates.  
Furthermore, the power law holds over 
the entire range in a double scaling limit.

This kind of voting model has been introduced 
in the literatures  of social-choice problems on 
preference formation in a voting population 
\cite{Berg,Berg2}. 
The voting paradox,
the possibility of individual preference patterns leading to
in-transitivity, ask about the likelihood that certain kinds of cycles occurs, 
given that
 people can choose at random among all possible profiles, rankings of choices.
In order that majority rule does work in decision making process,
or to fix the Condorcet's winner,  
 there must exist a transitive ordering among profiles. 
The voting model is a simple Polya-variety urn model.
A homogeneity parameter relates to measures of similarity among voters.
The model is a  rough model for contagion diseases, such that each 
occurrence increases the chance of further occurrences. 
We can interpret the homogeneity parameter as 
the contagion parameter or as the amount of
similarity-homogeneity among voters, the extent to which 
voters influence one another. 
It was concluded that as the preference similarity 
among voters increases, or stronger mutual
influence among voters, there is  a lesser chance for the paradox of
occurring.
Our conclusion is that in the ranking of the horses, the mutual 
influence among voters induces the scale invariance in the mixing.

The organization of this paper is as follows.
In  \S \ref{Model}, we introduce the voting model. 
We select a candidate (initial number of votes $s_{\mu}$) and show that 
the probability density function of the share of votes ,$u$, of 
the candidate obeys a gamma distribution function with the 
 shape exponent $s_{\mu}$ in the
thermodynamic limit $Z_{0}=N_{1}s_{1}+N_{0}s_{0} \to \infty$.
We also show that the joint probability density function 
of $u$ for any $k$ candidates 
is  given by the direct product of the 
gamma distributions in the same limit.
We discuss the scale invariance in the mixing of the binary candidates 
in \S \ref{Scale}.
The cumulative function $1-x_{\mu}$ of candidates $\bf{\mu}$
is  given by the incomplete gamma function.
The power-law relation
$1-x_{1}\sim (1-x_{0})^{\alpha}$ with the exponent
 $\alpha=s_{1}/s_{0}$ holds 
in the region $1-x_{0},1-x_{1} <<1$. 
Furthermore, in the double 
scaling limit $\{s_{\mu}\}\to 0$ and $Z_{0}\to \infty$ with 
$\alpha=s_{1}/s_{0}$ fixed, the relation 
$1-x_{1}=(1-x_{0})^{\alpha}$ holds exactly over the entire range
$0\le x_{0},x_{1} \le 1$.
Using the data on horse races, we verify these results
in \S \ref{Data}.  
We show that scale invariance holds over  the 
wide range of cumulative functions.
In addition, we show that the probability distribution functions 
of $u$ are well described  by gamma 
distributions.  
Section \ref{Conclusion} is dedicated to the summary and 
concluding remarks. 
Appendix A is devoted to the derivation of the joint probability
distribution function of $u$ for any $k$ candidates.
In Appendix B, we map the voting model to a branching process and 
easily derive the gamma distribution function.

\section{Voting Model for Binary Candidates}
\label{Model}

Consider a voting model for $N$ candidates.
Voters vote for them one by one, and the result of each voting 
is announced promptly. 
The time variable $t\in\{0,1,2\cdots,T\}$ 
counts the number of the votes.
The candidates are classified into two categories $\mu\in \{0,1\}$
and are called as binary candidates.
There are $N_{\mu}$  candidates in each category and 
$N_{0}+N_{1}=N$. The main result of this section is that
the scaled share of votes $u_{i}^{\mu}$ of a candidate $\mu$ obeys a 
gamma distribution with the shape exponent $s_{\mu}$ in 
the thermodynamic limit $N_{0},N_{1}\to \infty$. 

We denote the number of votes of $i$th candidate $\mu\in \{0,1\}$
at time $t$ as $\{X^{\mu}_{i,t}\}_{i\in \{1,\cdots,N_{\mu}\}}$.
At $t=0$, $X_{i,t}^{\mu}$ takes the initial value
$X_{i,0}^{\mu}=s_{\mu}>0$. If the $i$th candidate $\mu$ gets a vote at $t$,
$X_{i,t}^{\mu}$ increases by one unit.
\[
X_{i,t+1}^{\mu}=X_{i,t}^{\mu}+1. 
\]
A voter casts a vote for the total 
$N$ candidates at a rate proportional to $X_{i,t}^{\mu}$. 
The probability $P_{i,t}^{\mu}$ that the $i$th 
candidate $\mu$ gets a vote at $t$ is
\begin{eqnarray}
P_{i,t}^{\mu}&=&\frac{X^{\mu}_{i,t}}{Z_{t}} 
\\
Z_{t}&=&\sum_{\mu=0}^{1}
\sum_{i=1}^{N_{\mu}}X^{\mu}_{i,t}=N_{1}s_{1}+N_{0}s_{0}+t.
\end{eqnarray}

The problem of determining the 
probability of the $i$th candidate $\mu$ getting $n$ votes 
up to $T$ is equivalent to the 
famous P\'{o}lya's  urn problem 
\cite{Polya,Berg,Huillet,Hisakado}. 
If the change in $X_{i,t}^{\mu}$ is given by  
\[
\Delta X_{i,t}^{\mu}\equiv X_{i,t}^{\mu}-X_{i,t-1}^{\mu}, 
\]
the sequence $(\Delta X_{i,1}^{\mu},\cdots,
\Delta X_{i,T}^{\mu})$ is called  P\'{o}lya's urn sequence. 
This sequence is an exchangeable stochastic process, and 
the joint distribution of $(X_{i,1}^{\mu} 
\cdots,X_{i,T}^{\mu})$ is given by
\[
\mbox{Prob}(\Delta X_{i,1}^{\mu}=x_{1},\cdots,\Delta X_{i,T}^{\mu}=x_{T})
=\frac{(s_{\mu})_{k}(Z_{0}-s_{\mu})_{T-k}}{(Z_{0})_{T}}. 
\]
Here, $k=\sum_{t=1}^{T}x_{t}$ and
$(a)_{n}\equiv a\cdot (a+1)\cdot (a+2)\cdots (a+n-1)$ is 
the rising factorial. 
This distribution depends only on $k$, and not on the particular order
of $(x_{1},\cdots,x_{T})$. This distribution is invariant 
under the permutations of 
the entries and, hence, it is called exchangeable.

Furthermore, the expectation value of $\Delta X_{i,t}^{\mu}$, denoted by
$p_{\mu}$,  
does not depend on $t$.
\begin{equation}
p_{\mu} \equiv <\Delta X_{i,t}^{\mu}>=\frac{s_{\mu}}{Z_{0}}.
\end{equation}
The correlation function $\rho_{\mu}$ between $\Delta X_{i,t}^{\mu}$ 
and $\Delta X_{i,t'}^{\mu} \hspace*{0.2cm}
(t'\neq t)$ is also constant \cite{Hisakado}
as $\rho_{\mu}$.  
\begin{equation}
\rho_{\mu}\equiv \mbox{Corr}(\Delta X_{i,t}^{\mu},\Delta X_{i,t'}^{\mu})
\equiv \frac{<\Delta X_{i,t}^{\mu}\Delta X_{i,t'}^{\mu}>-p^{2}}{p(1-p)}
=\frac{1}{Z_{0}+1},\hspace*{0.3cm}t\neq t'  \label{Eq:corr}.
\end{equation}

The probability that the $i$th candidate $\mu$ gets $n$ votes up to $T$
is given by the beta binomial distribution
\begin{equation}
\mbox{Prob}(X_{i,T}^{\mu}-s_{\mu}=n)={}_{T}C_{n}\cdot 
\frac{(s_\mu)_{n}(Z_{0}-s_\mu)_{T-n}}{(Z_{0})_{T}}.
\end{equation}
$(a)_{n}$ is written as $(a)_{n}=\frac{\Gamma(a+n)}{\Gamma(a)}$ and
this relation can also be  written as
\begin{equation}
\mbox{Prob}(X_{i,T}^{\mu}-s_{\mu}=n)={}_{T}C_{n}\cdot 
\frac{\Gamma(s_{\mu}+n)}{\Gamma(s_{\mu})}
\frac{\Gamma(Z_{0}-s_{\mu}+T-n)}{\Gamma(Z_{0}-s_{\mu})}
\frac{\Gamma(Z_{0})}{\Gamma(Z_{0}+T)}.
\end{equation}
Using a definition of beta function 
$B(a,b)\equiv \frac{\Gamma(a)\Gamma(b)}{\Gamma(a+b)}$, we can rewrite
the expression as 
\begin{equation}
\mbox{Prob}(X_{i,T}^{\mu}-s_{\mu}=n)={}_{T}C_{n}\cdot 
\frac{B(s_{\mu}+n,Z_{0}-s_{\mu}+T-n)}{B(s_{\mu},Z_{0}-s_{\mu})}.
\end{equation}
$B(a,b)$ is also written as $B(a,b)=\int_{0}^{1}p^{a-1}(1-p)^{b-1}dp$, we 
get the next expression
\begin{equation}
\mbox{Prob}(X_{i,T}^{\mu}-s_{\mu}=n)={}_{T}C_{n}\cdot 
\int_{0}^{1}p^{n}(1-p)^{T-n}\frac{p^{s_{\mu}-1}(1-p)^{Z_{0}-s_{\mu}}}
{B(s_{\mu},Z_{0}-s_{\mu})}dp.
\end{equation}
After infinite counts of voting, i.e. $T\to \infty$, 
the share of votes 
$x_{i}^{\mu}\equiv \lim_{T\to \infty}
\frac{X_{i,T}^{\mu}-s_{\mu}}{T}$, 
becomes the beta distributed random variable
beta$(s_{\mu},Z_{0}-s_{\mu})$ on $[0,1]$.
\begin{equation}
p(x)\equiv \lim_{T\to \infty}
\mbox{Prob}(X_{i,T}^{\mu}-s_{\mu}=Tx)\cdot T=
\frac{x^{s_{\mu}-1}(1-x)^{Z_{0}-s_{\mu}-1}}{B(s_{\mu},Z_{0}-s_{\mu})}.
\end{equation}
Here, we use the identity 
$\lim_{T\to \infty}{}_{T}C_{Tx}p^{Tx}(1-p)^{T(1-x)}\cdot T=\delta(x-p)$.
This result has been derived by P\'{o}lya \cite{Polya}.

Next, we focus on the  thermodynamic limit $N_{0},N_{1} \to
\infty$ and $Z_{0}=N_{0}s_{0}+N_{1}s_{1}\to \infty$. 
The expectation value of $x_{i}^{\mu}$ is $<x_{i}^{\mu}>=p_{\mu}=
\frac{s_{\mu}}{Z_{0}}$. We introduce a variable $u_{i}^{\mu}\equiv
(Z_{0}-s_{\mu}-1)x_{i}^{\mu}$. 
The distribution function $p_{s_{\mu}}(u)$ in the  thermodynamic 
limit is given as
\begin{equation}
p_{s_{\mu}}(u)\equiv \lim_{Z_{0}\to \infty}p(x_{i}^{\mu}
=\frac{u}{Z_{0}-s_{\mu}-1})
=\frac{1}{\Gamma(s_{\mu})}e^{-u}u^{s_{\mu}-1}  \label{Eq:Gamma}.
\end{equation}
The share of votes, $u$, of a candidate $\mu$ 
obeys a gamma distribution function with $s_{\mu}$.

In general, the joint probability distribution 
function of the scaled share of votes of $k$ different candidates 
becomes the direct product of $k$ gamma distribution functions
in the limit $Z_{0}\to \infty$. 
We denote the $k$ candidates as $\{(\mu_{j},i_{j})\}_{j=1,\cdots,k}$
 and denote the scaled share of votes as 
$\{u_{j}\}_{j=1,\cdots,k}$. The joint probability distribution function 
is given as
\begin{equation}
p(u_{1},\cdots,u_{k})=\prod_{j=1}^{k}p_{s_{\mu_{j}}}(u_{j}).
\end{equation}
The derivation of the result is given in Appendix A. 
It should be noted that in the thermodynamic limit, 
the correlation among 
$\{u_{j}\}_{j=1,\cdots,k}$
vanishes. Hence, by mapping the voting problem to a 
continuous time branching 
process, we can derive the gamma
 distribution function $p_{s_{\mu}}(u)$ easily
(refer Appendix B).
In the branching process, the stochastic processes of the increase in 
$\{X_{i,t}^{\mu}\}$ are independent of each other.

\section{Scale Invariance in Mixing of Binary Candidates}
\label{Scale}

In this section, we discuss the mixing of the binary 
candidates. 
After many counts of voting $T\to \infty$, 
the binary candidates are distributed in the
space of $u$ according to the gamma distribution 
in the thermodynamic limit $Z_{0}\to \infty$.  If $s_{1}>s_{0}$,
 a candidate belonging to category $\mu=1$ has a higher 
probability of 
getting many votes than a candidate belonging to category $\mu=0$. 
Even the latter can obtain  
many votes. 
It is also possible that the former may 
get few votes. Thus, there is a mixing of the 
binary candidates. We see a scale invariant behaviour appears 
in the mixing. Between the cumulative functions of the binary candidates 
$1-x_{\mu}$, the power-law relation 
$1-x_{1} \sim (1-x_{0})^{\alpha}$ with the exponent $\alpha=s_{1}/s_{0}$
holds.

\begin{figure}[htbp]
\begin{center}
\includegraphics[width=10cm]{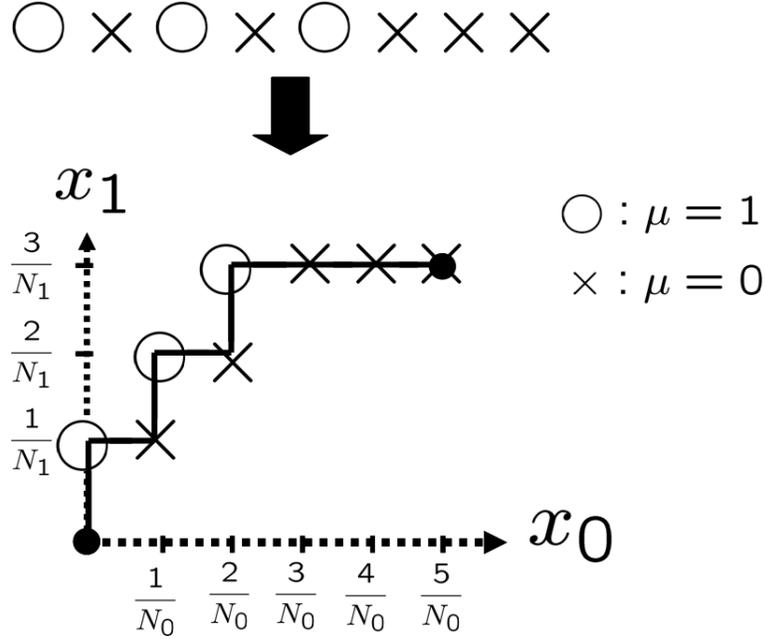}
\end{center}
\caption{ROC curve of mixing  configuration.
$\bigcirc$ represents candidate belonging to category $\mu=1$. 
$\times$ represents
candidate belonging to category $\mu=0$. At the top of the figure, 
the  order of three candidates from category $\mu=1$ and five candidates
 from category $\mu=0$ is shown.}
\label{Fig:path}
\end{figure}

In order to study the 
mixing configuration, we arrange the $N$ candidates 
according to the size of $u_{i}^{\mu}$ as 
\begin{equation}
u^{\mu_{1}}_{i_1}
> 
u^{\mu_{2}}_{i_2}
>\cdots
>u^{\mu_{N}}_{i_N},\hspace*{0.4cm}\mu_{k}\in \{0,1\}.
\end{equation}
Using the ranking information $\{\mu_{k}\}_{k=1,\cdots,N}$, we draw a 
path $\{(x_{0,k},x_{1,k})\}_{k=0,\cdots,N}$ in two-dimensional space 
$(x_{0},x_{1})$ from $(x_{0,0},x_{1,0})=(0,0)$ to
$(x_{0,N},x_{1,N})=(1,1)$ as
\begin{equation}
x_{\mu,k}=\frac{1}{N_{\mu}}\sum_{j=1}^{k}\delta_{\mu_j,\mu}.  
\end{equation}
See Fig. \ref{Fig:path}.
If $\mu_{k}=\mu$, the path extends  in
$x_{\mu}$ direction.  
The pictorial representation of the mixing of binary objects is known as
a receiver operating characteristic (ROC) curve \cite{Enleman}.
If $s_1>>s_0$, the binary candidates are well
separated on the axis of $u$,  and  the first
$N_{1}$ candidates belong to category  $\mu=1$ and the last $N_{0}$ 
candidates belong to category $\mu=0$.
The path goes straight from $(0,0)$ to $(0,1)$ and then turns right
to the end point $(1,1)$.  
If $s_1=s_0$, the path almost runs diagonally to the end point. 
If $s_1>s_0$ holds, 
the path resembles a upward convex curve from $(0,0)$ to $(1,1)$.

The distribution function of the candidate $\mu$ on the axis of $u$
is given by the gamma distribution  with the shape exponent $s_{\mu}$.
The ROC curve
$(x_{0}(t),x_{1}(t))$ of the  parameter $t\in [0,\infty]$ is
given by its cumulative function as
\begin{equation}
x_{\mu}(t)=\int_{t}^{\infty}p_{s_\mu}(u)du. 
\end{equation}
Using the incomplete gamma function of the first kind 
$\gamma(s,t)\equiv \int_{0}^{t}e^{-u}\cdot u^{s-1}du$ \cite{Abramowitz}, 
the ROC curve is given as
\begin{equation}
1-x_{\mu}(t)=\frac{1}{\Gamma(s_\mu)}\cdot \gamma(s_\mu,t). \label{eq:ROC}
\end{equation}
Near the end point, $(x_{0},x_{1})\simeq (1,1)$, in other words, in 
the small $u$ region $(t \simeq 0)$, the incomplete gamma
function $\gamma(s_\mu,t)$  behaves  as
\begin{equation}
\gamma(s_\mu,t)\sim  t^{s_\mu}.
\end{equation}
As $1-x_{s_\mu}(t)\propto t^{s_\mu}$, the following relation
holds:
\begin{equation}
1-x_{1}\sim (1-x_{0})^{\alpha}  \hspace*{0.4cm}
\mbox{with}\hspace*{0.4cm}
\alpha=\frac{s_1}{s_0}.      
\end{equation}
The density of good candidates, $\rho_{1}$, in terms of the cumulative 
function of bad candidates, $1-x_{0}$, is given as 
\begin{equation}
\rho_{1}=\frac{d(1-x_{1})}{d(1-x_{0})}\propto (1-x_{0})^{\alpha-1} .
\end{equation}
$\rho_{1}$ obeys  the power law with the exponent $\alpha-1$.

Furthermore, in the limit $(s_1,s_0)\to (0,0)$ with 
$\alpha=s_1/s_0$ fixed,
the relation $1-x_{1}=(1-x_{0})^{\alpha}$ holds.
The proof is given as follows. $\gamma(s,t)$ is 
expressed using Kummer's confluent hypergeometric function
$M(a,b,t)$  \cite{Abramowitz} as
\begin{equation}
\gamma(s,t)=\frac{1}{s}t^{s}\cdot M(s,s+1,-t). 
\end{equation}
The cumulative function $1-x_{\mu}(t)$ is then given as
\begin{equation}
1-x_{\mu}(t)=\frac{t^{s_\mu}}{\Gamma(s_\mu+1)}\cdot M(s_\mu,s_\mu+1,-t)
\label{eq:power_gamma}.
\end{equation}
Thus, we obtain
\begin{equation}
(1-x_{0})^{\alpha}=(1-x_{1})
\frac{\Gamma(s_1+1)}{M(s_1,s_1+1,-t)}
\left(\frac{M(s_0,s_0+1,-t)}{\Gamma(s_0+1)}\right)^{\alpha}.
\end{equation}
In the limit $s_\mu\to 0$, both $\Gamma(s_\mu+1)$ and 
$M(s_\mu,s_\mu+1,-t)$ become equal to $1$
and the following relation holds.
\begin{equation}
1-x_{1}=(1-x_{0})^{\alpha}\hspace*{0.2cm},\hspace*{0.2cm}0\le x_{0},x_{1}\le 1 \label{eq:scale}.
\end{equation}
Thus, the scale-invariant relation holds over the entire
range $0\le x_{0},x_{1} \le 1$. The feature is remarkable from the
viewpoint of statistical physics. Usually, 
the power-law relation does hold only in the tail.

 The relative probability that  a candidate gets
 the first vote $(t=0)$ is given by $s_\mu$. If 
the  candidate get the first vote,
 his/her score increases by 1 and the relative probability becomes $s_\mu+1$. 
In the limit $s_\mu \to 0$,
the additional score $+1$  or the weight
of a single vote  becomes crucially important. 
The probability that the candidate gets the next vote
 becomes equal to 1, which is exemplified by the behaviour 
of $\rho_{\mu}$, given by  eq.(\ref{Eq:corr}).
\begin{equation}
\rho_{\mu}=\frac{1}{Z_{0}+1}
=\frac{1}{N_{0}s_{0}+N_{1}s_{1}+1}\to 1 \hspace*{0.3cm}
\mbox{if}\hspace*{0.2cm} \{s_{\mu}\}\to 0. 
\end{equation}
After infinite counts of voting, the candidate occupies the first
position in the order of candidates according to the number of votes. 
Then, we neglect this candidates in the voting problem and consider 
 the remaining 
$N-1$ candidates. 
Similarly, if a candidate is selected randomly with the relative 
probability $s_\mu$, he/she occupies the second position.
Thus, the voting problem reduces to  a random choice problem with the  
relative probability $s_\mu$ in the limit $\{s_\mu \}\to 0$.
At $(x_{0},x_{1})$ on the ROC curve, the probability that 
the next candidate belongs category 
$\mu$ is proportional to $(1-x_{\mu})s_\mu \hspace*{0.1cm}$.  
The coordinates of the ROC curve $(x_{0},x_{1})$ 
grow according to the following relation:
\[
dx_{\mu}\propto (1-x_{\mu})\cdot s_\mu. 
\]       
Solving this relation, we get eq.(\ref{eq:scale}).

Finally, we discuss the limit in the derivation of the exact scale
invariance. 
In the derivation
of the gamma distribution, we take the thermodynamic limit 
$Z_{0}=N_{1}s_{1}+N_{0}s_{0}\to \infty$. With the gamma distribution, 
eq.(\ref{eq:scale}) holds in the limit $\{s_{\mu}\}\to 0$.  
In order that eq.(\ref{eq:scale}) holds, these two limits, $Z_{0}\to
\infty$
and $\{s_{\mu}\}\to 0$, should go together. 
$\{s_{\mu}\}$ approaches zero more slowly than 
$\{N_{\mu}\}$ approaches infinity. In other words, in the double scaling 
limit $Z_{0}\to \infty$ and $\{s_{\mu}\} \to 0$ with $\alpha=s_{1}/s_{0}$
fixed, eq.(\ref{eq:scale}) holds. So the above intuitive explanation 
of the exact scale invariance may be
too naive. If we take the limit $\{s_{\mu}\}\to 0$ without 
the limit $Z_{0}\to \infty$, $\rho_{\mu}$ becomes 1.
The firstly chosen candidate get all the remaining votes and there does
not occur the mixing of the binary candidates. The double scaling limit
is  crucial in the emergence of the exact scale invariance.

\section{Data Analysis of Horse Races}
\label{Data}  

We verify the results of the voting model, particularly  the 
scale invariance in the mixing of binary candidates.
We study all the data  on horse race betting obtained from the  
Japan Racing Association (JRA) for the period  1986 to 2006. 
There have been 71549 races
and in which a total of  901366 horses have participated. 
We select
the winning horses as  candidate belonging to category $\mu=1$.
For candidate belonging to category $\mu=0$, we consider two cases;
losing horses and horses finishing second. 
 In a race, no one knows which horse will win. 
 Betters only have partial information
on the horses, which is embedded in the initial values 
$\{s_{\mu}\}$. 
The results of betting are
announced at short intervals. 
Betters usually presume that 
the horses which get many votes 
are strong. 
They come to know which horses are considered to be strong 
by other betters.
These features are 
incorporated in the voting model.
Betters do not always bet to  strong horses.
Some betters may prefer betting to a horse that can coin more 
money
even if it is considered to be `weaker' than a horse that can coin 
less money. 
However, in the bet to win, only the better who bets to the winning horse
coin the bet. Hence, the assumption 
is not so unrealistic. We also note the reason why we can 
treat multiple categories, 2nd finishing horse and losing horse, as the category $\mu=0$. 
For the betters, the only difference between the losing horses 
and finishing second ones is their confidence. 
By tuning parameter $s_{0}$, we can treat the two categories on the same footing.
  
Next, we explain the meaning of the initial values $\{s_{\mu}\}$.
The probability that a candidate $\mu$ is selected
is proportional to $s_{\mu}$ as
$<\Delta X_{i,t}^{\mu}>=s_{\mu}/Z_{0}$. The ratio $s_{1}/s_{0}$
is a measure of  the accuracy of the knowledge of betters.
On the other hand, $\rho_{\mu}$ is given 
by eq.(\ref{Eq:corr}). If the scale of $\{s_{\mu}\}$ is small, 
the decisions of betters are crucially affected by the choices
of other betters.
In the limit $\{s_{\mu}\}\to \infty$, their decisions are not
affected by the choices of other betters. 
The scale of $\{s_{\mu}\}$ is a measure of  the 
degree of similarity ('copycat') of the choices of betters.
  
In the early stage of voting, $\{s_\mu\}$ is 
the only available information.
Voters decide on horses on the basis of $\{s_{\mu}\}$ and they are  
`intelligent', because their decisions  
are not affected by the choices of other betters.   
As the voting process proceeds, the
importance of the cumulative number of votes exceeds that of the initial 
scores, and 
voters become `copycat'.
 If one control 
the scale of $\{s_\mu\}$ (or the  weight of a single vote), 
the passage timing from the initial `intelligent'  
stage to the late `copycat' stage  should change.

\begin{table}[htbp]
\caption{\label{tab:data0}
Data on horse race betting obtained from the Japan Racing Association
 (JRA) for the period 1986 to 2006. There are 71549 races and 71650
 winning (finishing first) horses. 71590 horses are finishing second.
The difference between $N^{Win}$ and $N^{2nd}$ indicates that there
 occurs a tie in the race. In the third column, we show the  average
 value of the share of the votes in each category. The fourth column 
shows the values $v^{\nu}/c$, here $c$ is the estimated value of 
the scale parameter in (\ref{eq:c}). About the estimation of $c$, please 
see the main text and Figure \ref{fig:P_v}.}
\begin{tabular}{cccc}
Category $\nu$ & $N^{\nu}$ &  
$v^{\nu}\hspace*{0.1cm}  [\%]$  & $v^{\nu}/c$   \\
\hline
Win  & 71650   & 21.23& 1.769  \\
2nd  & 71590   & 15.40 & 1.283  \\
Lose  & 829716 & 6.80& 0.567 \\
\end{tabular}
\end{table}

We denote the three categories of horses 
as $\nu \in \{$Win, 2nd, Lose$\}$ and 
the number of horses in each category as $N^{\nu}$.
$v_{i}^{\nu}$
 denotes the share of votes of the $i$th horse in the category $\nu$, 
 and  
 $v^{\nu}$
 denotes   the average value of $v_{i}^{\nu}$.
In Table \ref{tab:data0}, we summarize the data on horse races.
A difference between $N^{Win}$ and $N^{2nd}$ indicates that there is a tie
in the race.

\begin{figure}[htbp]
\begin{center}
\includegraphics[width=12cm]{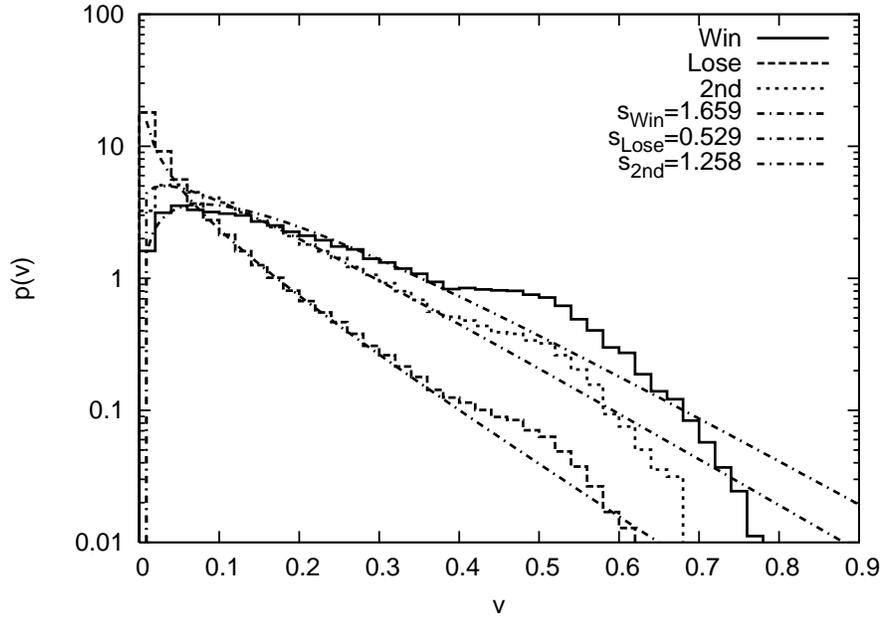}
\end{center}
\caption{Logarithmic plot of probability 
distribution functions $p(v)$ of shares of votes.
The curves from the top to bottom indicate the data for
 $\nu=$Win (solid), 2nd (dashed) 
 and
Lose (dotted).
The gamma distribution functions 
with $c=0.12$ and $s_{\nu}$ are also plotted (chain lines).}
\label{fig:P_v}
\end{figure}

We have shown that the share of votes, $u$, obeys a 
gamma distribution function with $s_{\mu}$.
In order to check whether $v_{i}^{\nu}$ obeys a 
 gamma distribution function, 
we have to set the scale $c$ between $v_{i}^{\nu}$ and 
$u$ as follows:
\begin{equation}
v_{i}^{\nu}=c \cdot u.  \label{eq:c}
\end{equation}
The same $c$ should be used for all categories.
Assuming that $u$ obeys the gamma probability 
distribution with $s_\nu$,
 $v_{i}^{\nu}$ obeys the following probability distribution function: 
\[
p(v_{i}^{\nu}=v)=p_{s_\nu}(v)=
\frac{1}{c\cdot \Gamma(s_\nu)}(\frac{v}{c})^{s_\nu-1}\exp(-\frac{v}{c}).
\]
The expectation value of $v_{i}^{\nu}$ is 
\[
<v_{i}^{\nu}>_{\mu}= \int_{0}^{\infty}p_{s_\nu}(v)v dv
=c\cdot s_\nu. 
\]
If we set $c$, it is possible to estimate $s_{\nu}$ 
of the horses in category $\nu$ as $s_{\nu}=v^{\nu}/c$.

Figure \ref{fig:P_v} shows the probability distribution 
functions $p(v)$ of $v_{i}^{\mu}$. In the same figure, we show the 
result of fitting with the gamma probability functions.
Using the least square method in the range $v \in[0.01,1.0]$, 
we set $c=0.12$ and 
$s_{Win}=1.659,s_{2nd}=1.258$ and 
$s_{Lose}=0.529$. 
Comparing with the values in the fourth column in  
Table \ref{tab:data0}, it is observed the values of $s_{\nu}$ 
and $v^{\nu}/c$ are close to each other in all categories, implying
that the bulk shapes of the probability functions of 
$v_{i}^{\nu}$ are well described by the gamma distributions.
We also notice clear discrepancies in the figure. $p(v)$ 
 does not obey the gamma 
distribution for the larger shares.  The bulk shape of $p(v)$ 
is not crucial in our argument, because we are
interested in the critical properties, or small win bet fraction
regime. We think the discrepancies come from that the voters' confidence 
$s_{\mu}$ has some variance.

\begin{figure}[htbp]
\begin{center}
\includegraphics[width=12cm]{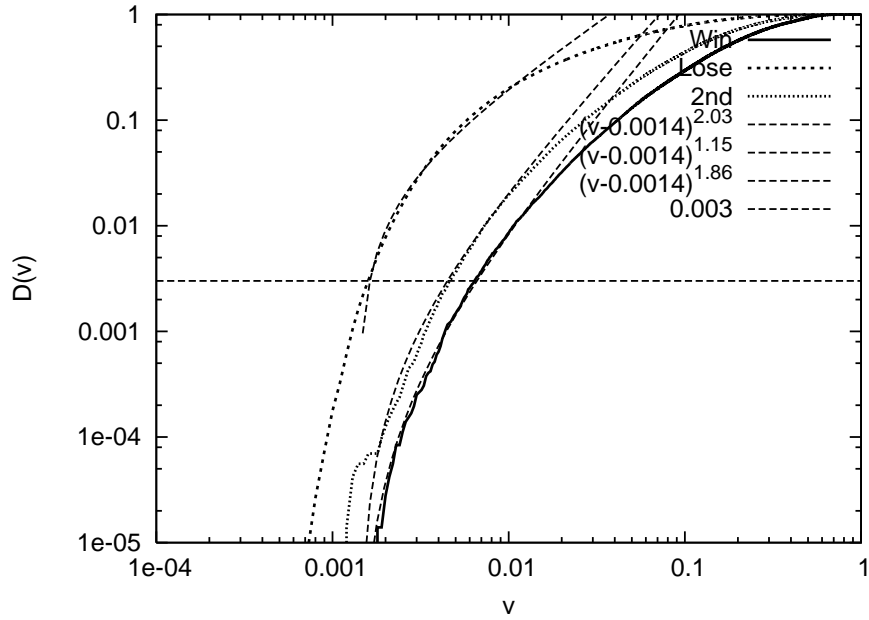}
\end{center}
\caption{Double-logarithmic plot of cumulative  
distribution functions $D(v)$ of shares of votes.
The curves from the top to bottom indicate the data for $\nu=$ Lose,2nd and
Win. The fitted functions $(v-v_{c})^{s'_{\mu}}$ are also plotted.
We set $v_{c}=0.0014$.}
\label{fig:D_v}
\end{figure}

We study the cumulative functions $1-x_{\mu}$ in the small share region,
$v\to 0$. Figure \ref{fig:D_v} shows the cumulative functions $D(v)$ of 
$v_{i}^{\mu}$, which is defined as
\begin{equation}
D(v)\equiv \int_{0}^{v}p(v)dv. 
\end{equation}
We are interested in the power law behaviour of $D(v) \propto
v^{s_{\mu}}$  and the figure 
shows the double logarithmic plot. We see that they do not obey the power
law, as have been predicted in eq.(\ref{eq:power_gamma}).
 In the figure, we show the result of fitting result with 
$(v-v_{c})^{s'_{\mu}}$. We set $v_{c}=0.0014$ and the figure shows that
the winning and finishing second horses's $D(v)$ obey the power law with 
cut-off $v_{c}$. On the other hand, about the losing horses, the fitting
only works for the region $D(v)\ge D_{c}\equiv 0.003$ and $v \ge v_{c}$.
The reason why $D(v)$ does not obey the power law is not clear. We
think that there are  some voters who want to vote to the horses 
with remarkable small shares. The odds are very large and for 
the voters, the horses look very attractive. If so, we can understand
the existence of the cut-off $v_{c}$.

\begin{figure}[htbp]
\begin{center}
\includegraphics[width=12cm]{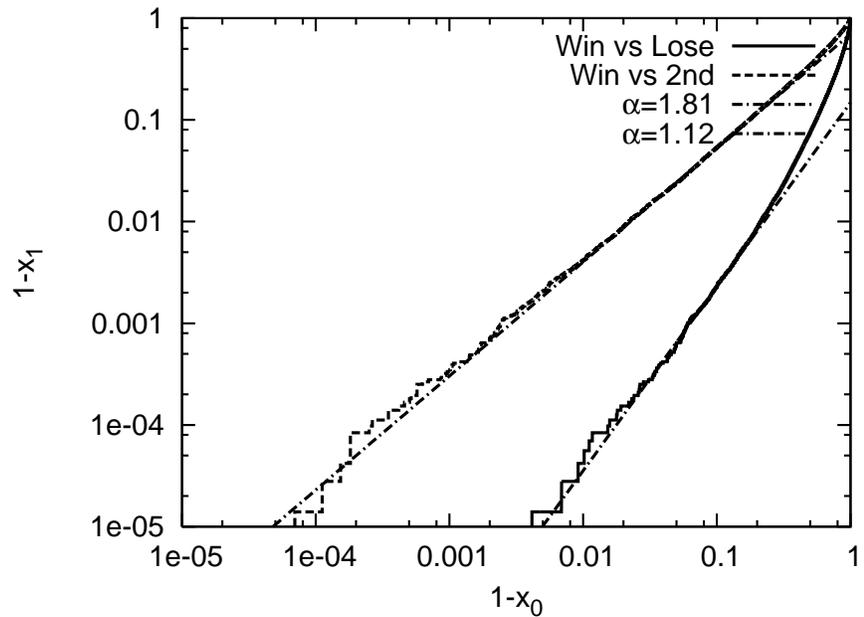}
\end{center}
\caption{Double-logarithmic plot of ROC curves $(1-x_{0},1-x_{1})$.
The curves of the Win-Lose pair (solid line) and the Win-2nd pair 
(dashed line) are plotted. 
The fitting curves given by 
$1-x_{1}=a\cdot (1-x_{0})^{\alpha}$ (dash-dotted line) are also plotted.}
\label{fig:ROC_curve}
\end{figure}

\begin{table}[htbp]
\caption{\label{tab:data}
The initial value $s_{\mu},s'_{\mu}$ in each category and the critical 
exponent $\alpha$ are plotted. In the last two column, we show the 
predicted values of $\alpha$ by the voting model.
}
\begin{tabular}{cccccccc}
Pair & $s_1$ & $s_0$ & $s'_1$ &$s'_0$ & $\alpha$ & $s_1/s_0$ & $s'_1/s'_0$\\
\hline
Win vs Lose & 1.659 & 0.529 & 2.03 &  1.15 & 1.81 &  3.134 & 1.765 \\
Win vs 2nd & 1.659 & 1.258 & 2.03 & 1.86 & 1.12   &    1.318 &  1.091 \\
\end{tabular}
\end{table}

We study the mixing properties
of the binary horses by employing the method explained in the 
text. We adopt the Win-Lose pair 
and Win-2nd pair as the binary pairs.
Figure \ref{fig:ROC_curve} shows the double-logarithmic plot 
of the ROC curve $(1-x_{0},1-x_{1})$ for the two pairs.
The plots show scale-invariant behaviour over the wide
range of $1-x_{1}$. In the case of 
the Win-2nd pair, scale invariance holds 
over the range $10^{-5}<1-x_{1}
<10^{-1}$,  which can be anticipated from the bahaviour
of $D(v)$. About the Win-Lose pair, the range 
is restricted for $1-x_{0}\ge D_{c}=0.003$. In order to see 
the scale invariance for the region $1-x_{0}\le D_{c}$, many more
results of the races ($N^{Win} \sim 10^{6}$) are necessary.  
Using the least square method in the range $0\le 1-x_{0}\le 
0.1$, we estimate the critical exponent $\alpha$. 
The values of the parameters 
and other data are summarized in Table \ref{tab:data}.
The estimated values of $\alpha$ 
are considerably different from those 
predicted by the model; i.e. $s_1/s_0$.
In the table, we also show the values $s'_{\mu}$ estimated by fitting
$D(v)$ with $(v-v_{c})^{s'_{\mu}}$. The estimated values of $\alpha$
 are closer than the values from the bulk values $s_{\mu}$.

\section{Concluding Remarks}
\label{Conclusion}

In this study, we have introduced a simple voting model in order to
discuss the mixing of binary candidates with initial number of votes  
$s_0$ and $s_1$. 
As the voting process proceeds, the candidates are mixed in
the space of the share of votes, $u$.
We have shown that the probability distribution of $u$ of a  
candidate $\mu$ obeys a gamma distribution function  
with the shape exponent $s_{\mu}$ in the thermodynamic limit $Z_{0} \to 0$.
The joint probability distribution of $k$ different candidates is
given as the direct product of the gamma distributions.
The mixing configuration of the binary candidates 
exhibits   scale invariance in the small $u$ region.
In particular, in the double scaling limit $Z_{0}\to \infty$ and 
$\{s_{\mu}\}\to 0$ with $\alpha=s_{1}/s_{0}$ fixed,
the scale invariance holds over the entire range.  
The cumulative function of the binary candidates
obeys $1-x_{1}=(1-x_{0})^{\alpha}$ for $0\le x_{0},x_{1} \le 1$.

\begin{figure}[h]
\begin{center}
\includegraphics[width=12cm]{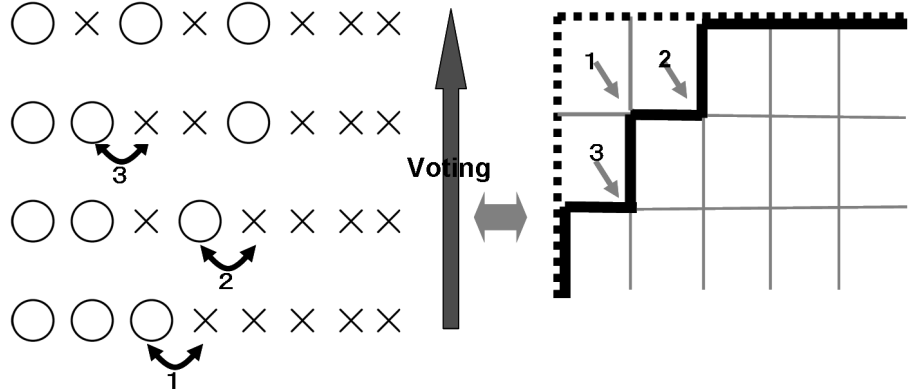}
\end{center}
\caption{Voting model and Random Young diagram model.
As the voting process proceeds, the order of the binary candidates
and the Young diagram change.
The complementary space of the ROC curve corresponds the 
Young diagram. }
\label{fig:Young}
\end{figure}

The data on horse races obtained from  JRA 
also show that scale invariance holds over 
the wide range of cumulative functions.
The distribution functions of the share of votes, $u$, are 
to some extent 
described by the gamma distribution functions, implying
that the behaviour of betters is described by the voting model.
However a clear discrepancy is observed in the critical behaviour.
The bulk properties of the probability function $p(v)$ and
the
critical properties of the cumulative functions $D(v)$ should be
discussed separately.
Although our voting model describes the mechanism of scale invariance in 
the mixing of binary candidates, it may be too simple to describe 
the behaviour of betters in real cases.
Thus far, dividends have been reported to exhibit 
power-law behaviour. 
Another betting model
has been proposed in \cite{Park,Hisakado2}.
A detailed study of real data, in particular the  time series
of the number of votes, should clarify the mechanism of scale invariance 
in betting systems \cite{Mori2}.

We also note that our model is related to the random Young diagram
problem \cite{Andrews}. This problem pertains to the 
probabilistic growth of 
a Young diagram. 
A parabolic shape \cite{Rost}
and a quadrant shape \cite{Jockusch} have been obtained for the 
asymptotic shape. The complementary part of the ROC curve, which is
embedded in the fourth quadrant, corresponds the Young diagram.
In our model, the ROC curve $(x_{0}(t),x_{1}(t))$ 
given by (\ref{eq:ROC})
describes the asymptotic shape of the Young diagram. 
 In particular, it is described by 
the relation $1-x_{1}=(1-x_{0})^{\alpha}$ 
 in the double scaling limit. 
Figure \ref{fig:Young} shows the correspondence between the voting
model and the random Young diagram problem. As the  voting process
proceeds, the 
order of the binary candidates and the Young diagram 
change.

\begin{figure}[t]
\begin{center}
\includegraphics[width=13cm]{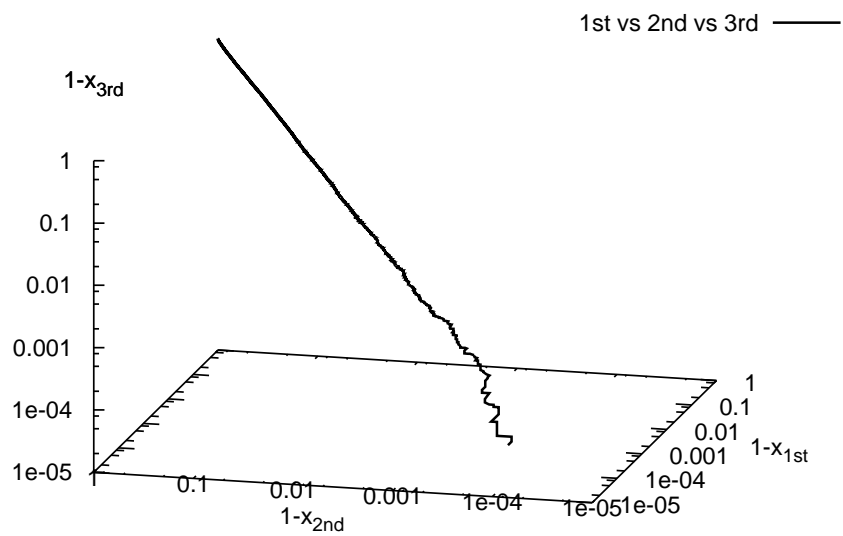}
\end{center}
\caption{Triple-logarithmic plot of ROC curve 
$(1-x_{1st},1-x_{2nd},1-x_{3rd})$. $x_{\nu}$ denotes 
the cumulative function of the horses finishing 
1st, 2nd and 3rd.
}
\label{fig:ROC_curve_2}
\end{figure}

It is also possible to study the voting model with 
many categories of candidates with the usage of many different 
initial values $\{s_{\mu}\}$. $u$
 of the candidates in each category
becomes a gamma distributed random variable.
 Scale invariance
 does hold between any pair of categories.   
Figure \ref{fig:ROC_curve_2} shows the triple-logarithmic 
plot of the cumulative functions of the winning ($x_{1st}$), 
finishing second ($x_{2nd}$) and finishing third 
($x_{3rd}$) horses. 
In the linear part of the curve, scale invariance holds between any
pair of categories.

\section*{Acknowledgment}
We thank Dr. Emmanuel Guitter 
for his useful discussions on the  branching process. 
We also  thank Takumi Nakaso for helping us  with 
the data analysis of horse races. This work was supported by
Grant-in-Aid for Challenging Exploratory Research 21654054.

\appendix

\section{Joint probability distribution function}

We start from the expression of the joint probability 
function given by
\begin{equation}
\mbox{Prob}(\{X_{i_{j},T}^{\mu_{j}}-s_{\mu_{j}}=n_{j}\}_{j=1,\cdots,k})
=\frac{T!}{(Z_{0})_{T}}\prod_{j=1}^{k+1}
\left[\frac{(s_{\mu_{j}})_{n_{j}}}{n_{j}!}\right].
\end{equation}
Here, $s_{\mu_{k+1}}\equiv Z_{0}-\sum_{j=1}^{k}s_{\mu_{j}}$
and $n_{k+1}\equiv T-\sum_{j=1}^{k}n_{j}$. 
Using the Dirichlet 
distribution function, we can rewrite the expression as
\begin{equation}
\mbox{Prob}(\{X_{i_{j},T}^{\mu_{j}}-s_{\mu_{j}}=n_{j}\}_{j=1,\cdots,k}) 
=\frac{T!}{\prod_{i=1}^{k+1}n_{i}!}\prod_{i=1}^{k}
\left[\int_{0}^{1-\sum_{j=1}^{i-1}p_{j}}dp_{i}
\right]\prod_{i=1}^{k+1}\left[\frac{p^{n_{i}+s_{\mu_{i}}-1}}
{\Gamma(s_{\mu_{i}})}\right]\Gamma(Z_{0}).
\end{equation}
The expectation value of $\Delta X_{i_{j},t}^{\mu_j}
=X_{i_{j},t+1}^{\mu_j}-X_{i_{j},t}^{\mu_j}$ is given by
\begin{equation}
p_{\mu_j}=<\Delta X_{i_{j},t}^{\mu_j}>=\frac{s_{\mu_j}}{Z_{0}}.
\end{equation}
The correlation between $\Delta X_{i_{j},t}^{\mu_j}$ and $\Delta
X_{i_{k},t}^{\mu_k} \hspace*{0.2cm}(k\neq j)$ is given as
\begin{equation}
\rho_{\mu_j,\mu_k}=-\sqrt{\frac{s_{\mu_j}s_{\mu_k}}{
(1-\frac{s_{\mu_j}}{Z_{0}})
(1-\frac{s_{\mu_k}}{Z_{0}})}}\frac{1}{Z_{0}(1+Z_{0})}.
\end{equation}

By changing the integral variables from $\{p_{i}\}_{i=1,\cdots,k}$ 
to $\{h_{i}\}_{i=1,\cdots,k}$ as
$p_{i}=(1-\sum_{j=1}^{i-1}p_{j})h_{i}=\prod_{j=1}^{i-1}(1-h_{j})h_{i}$,
we obtain
\begin{eqnarray}
&&\mbox{Prob}(\{X_{i_{j},T}^{\mu_{j}}-s_{\mu_{j}}=n_{j}\}_{j=1,\cdots,k}) 
\nonumber \\
&=&\frac{\Gamma(Z_{0})}{\Gamma(s_{\mu_{k+1}})}
\prod\left[
\frac{1}{\Gamma(s_{\mu_{i}})}\left(
{}_{T-\sum_{j=1}^{i-1}n_{j}}C_{n_{i}}\cdot
\int_{0}^{1}dh_{i}
h_{i}^{n_{i}+s_{\mu_{i}}-1}(1-h)^{T-\sum_{j=1}^{i}n_{j}+Z_{0}-\sum_{j=1}^{i}s_{\mu_{j}}-1} 
\right)
\right] . \nonumber \\ 
\end{eqnarray}
We focus on the share of votes of candidates in the limit $T\to \infty$.
We introduce  $y_{i}$ as $n_{i}=(T-\sum_{j=1}^{i-1}n_{j})y_{i}
=T\prod_{j=1}^{i-1}(1-y_{j})y_{i}$ and 
 define the joint distribution function as
\begin{equation}
P(\{y_{j}\}_{j=1,\cdots k})\equiv \lim_{T\to \infty}
\mbox{Prob.}(\{X_{i_{j},T}^{\mu_{j}}-s_{\mu_{j}}=
T\prod_{l=1}^{j-1}(1-y_{l})y_{j} \}_{j=1\cdots k})
\cdot \prod_{i=1}^{k}(T-\sum_{j=1}^{i-1}n_{j}).
\end{equation}
The joint function $P(\{y_{j}\}_{j=1,\cdots,k})$ is given by
\begin{equation} 
P(\{y_{j}\}_{j=1,\cdots,k})=\frac{\Gamma(Z_{0})}
{\Gamma(s_{\mu_{k+1}})}\prod_{i=1}^{k}\left[
\frac{1}{\Gamma(s_{\mu_{i}})}y_{i}^{s_{\mu_{i}}-1}
(1-y_{i})^{Z_{0}-\sum_{j=1}^{i}s_{\mu_{j}}-1}
\right].
\end{equation}
We introduce the variable $x_{i}$ as
$x_{i}=(1-\sum_{j=1}^{i-1}x_{j})y_{i}$, which is related to $n_{i}$
as $n_{i}=T\cdot x_{i}$. The joint probability function 
$P(\{x_{j}\}_{j=1,\cdots,k})$ is then given as
\begin{equation}
P(\{x_{j}\}_{j=1,\cdots,k})=\frac{\Gamma(Z_{0})}{\Gamma(s_{\mu_{k+1}})}
\prod_{i=1}^{k} \left[
\frac{1}{\Gamma(s_{\mu_{i}})}x_{i}^{s_{\mu_{i}}-1} \right]
(1-\sum_{j=1}^{k}x_{j})^{s_{\mu_{k+1}}-1}.
\end{equation}
Finally, we introduce the variable $\{u_{i}\}$ as
$u_{i}\equiv (s_{\mu_{k+1}}-1)x_{i}$. In the thermodynamic limit
$Z_{0},s_{\mu_{k+1}}\to \infty$, we obtain
\begin{equation}
P(\{u_{j}\}_{j=1,\cdots,k})=\prod_{j=1}^{k}
\left[
\frac{e^{-u_{j}}}{\Gamma(s_{\mu_{j}})}
u_{j}^{s_{\mu_{j}}-1}
\right]=\prod_{j=1}^{k} p_{s_{\mu_{j}}}(u_{j}).
\end{equation}

\section{Continuous time branching process}

\begin{figure}[htbp]
\begin{center}
\includegraphics[width=15cm]{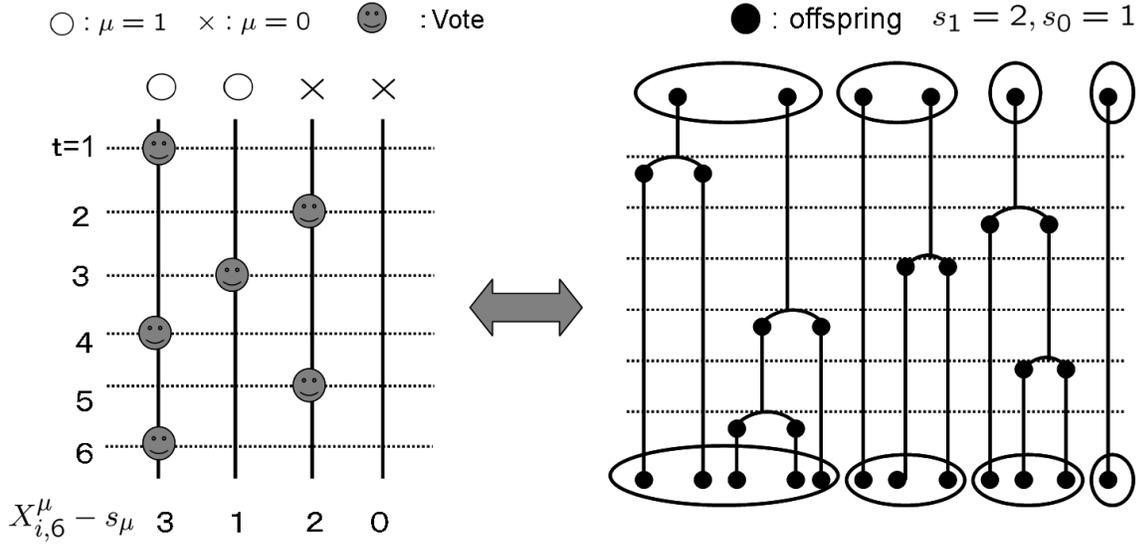}
\end{center}
\caption{Mapping voting model to branching process. The left-hand side
 figure  shows a voting process
with $N_{1}=N_{0}=2$. 
$\bigcirc$ represents candidate belonging to category $\mu=1,s_{1}=2$ and 
$\times$ represents 
candidate belonging to  category$\mu=0,s_{0}=1$.
The right-hand side  figure shows the corresponding branching process.
$\bullet$ represents the initial individual and offspring.
Candidate belonging to category $\mu=1 (\bf{0})$ is 
composed of two individuals (one individual).  
}
\label{Fig:mapping}
\end{figure}

We translate the discrete time voting problem
$\{X_{i,t}^{\mu}\}_{i=1,\cdots,
N_{\mu}}$ to a continuous time  
branching process $\{X_{i}^{\mu}(t)\}_{i=1,\cdots,N_{\mu}}$ 
\cite{Haccou}, because 
the latter is more tractable than the former \cite{David}.
Figure \ref{Fig:mapping} shows the mapping process.
Let $X_{i}^{\mu}(t)$
denote the number of offspring of $s_\mu$ individuals.  
Each individual is substituted by two offspring at its death (branching) and
the probability that an individual dies during time $dt$ is given by $dt$.
The number of offspring of each individual is denoted as
$\{x_{i,k}^{\mu}(t)\}_{k=1,\cdots,s_\mu}$. 
\begin{equation}
X_{i}^{\mu}(t)=\sum_{k=1}^{s_\mu}x_{i,k}^{\mu}(t) \hspace*{0.5cm}
,\hspace*{0.5cm}x_{i,k}^{\mu}(0)=1.
\end{equation}
The substitution of the individuals by two offspring 
corresponds to the process of getting a vote. 
The frequency of deaths or  the
probability of getting another vote is proportional to $X_{i}^{\mu}(t)$.
This relation is the same as that in the discrete time voting model.
The counts  of voting,$t$, corresponds to the counts of 
branchings. If branching takes place $t$ times up to $t'$,
the following relation holds.   
\[
X_{i}^{\mu}(t')=X_{i,t}^{\mu}
\]

The expectation values $<x_{i,k}^{\mu}(t)>$ and $<X_{i}^{\mu}(t)>$ 
increase with $e^{t}$. Next,  
 we introduce
the  scaled variables $U_{i}^{\mu}(t)$ and $u_{i,k}^{\mu}(t)$ as
\begin{equation}
U_{i}^{\mu}(t)\equiv e^{-t}X^{\mu}_{i}(t) \hspace*{0.3cm}
\mbox{and}\hspace*{0.3cm}
u_{i,k}^{\mu}(t)\equiv e^{-t}x_{i,k}^{\mu}(t).
\end{equation}
We focus on the following probability distributions:
\begin{eqnarray}
p_{s_\mu}(u)du &\equiv & \lim_{t\to \infty}
\mbox{Prob}(u\le U_{i}^{\mu}(t)\le u+du) \\
p(u)du &\equiv & \lim_{t\to \infty}
\mbox{Prob}(u\le u_{i,k}^{\mu}(t)\le u+du). 
\end{eqnarray}

\begin{figure}[htbp]
\begin{center}
\includegraphics[width=6cm]{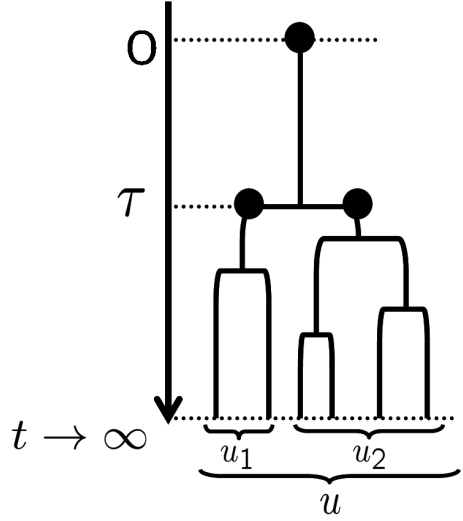}
\end{center}
\caption{Pictorial representation of 
self-consistent relation among $u,u_{1}$ and $u_{2}$.
An individual splits at $t=\tau$ for the first time producing  
two offspring
appears.Because of the time lag $\tau$, the relation 
$u=(u_{1}+u_{2})e^{-\tau}$ holds.}
\label{Fig:u2u}
\end{figure}

In order to obtain $p(u)$, we consider the situation in which  
an individual splits at $t=\tau$ for the first time. The 
resulting two offspring continue the branching process.
The scaled number of offspring of the individual is denoted as $u$.
Those of the two offspring are denoted as 
$u_{1}$ and $u_{2}$. Figure \ref{Fig:u2u} gives a pictorial
representation of the relation among $u,u_{1}$ and $u_{2}$.
We observe that these variables satisfy the following relation:
\[
u=(u_{1}+u_{2}) e^{-\tau}.
\]
Furthermore, $u_{1}$ and $u_{2}$ obey the same probability distribution
 as that obeyed by $u$,  and the probability that an individual 
splits for the first time 
during $\tau \le t \le \tau+d\tau$ is $e^{-\tau}d\tau$. Thus, we obtain 
\begin{equation}
p(u)=\int_{0}^{\infty}e^{-\tau}d\tau\int_{0}^{\infty}du_{1}
\int_{0}^{\infty}du_{2}p(u_{1})p(u_{2})\delta(u-(u_{1}+u_{2})e^{-\tau}).
\end{equation}
Introducing $X=e^{-\tau}$, the relation is rewritten as
\begin{equation}
p(u)=\int_{0}^{1}dX\int_{0}^{\infty}du_{1}
\int_{0}^{\infty}du_{2}p(u_{1})p(u_{2})\delta(u-(u_{1}+u_{2})X).
\end{equation}
Using the Laplace transform of $p(u)$, $\hat{p}(s)\equiv 
\int_{0}^{\infty}p(u)e^{-su}du$, it can be shown that 
$\hat{p}(s)$ satisfies the following integral equation:
\begin{equation}
\hat{p}(s)=\frac{1}{s}\int_{0}^{s}\hat{p}(v)dv \label{eq:integral}.
\end{equation}
Differentiating (\ref{eq:integral}) with respect to  $s$, 
we obtain the following differential equation.
\begin{equation}
s\frac{d \hat{p}(s)}{ds}=\hat{p}^{2}(s)-\hat{p}(s) \label{eq:diff}.
\end{equation}
(\ref{eq:diff}) can be solved easily to obtain
\begin{equation}
\hat{p}(s)=\frac{1}{1+as}.
\end{equation}
Using the normalization condition $<u>=1$ and
the inverse Laplace transform, we get 
\begin{equation}
p(u)=e^{-u}.
\end{equation}
We obtain $p_{s_{\mu}}(u)$ by convolution as
\begin{eqnarray}
p_{s_{\mu}}&=&\prod_{i=1}^{s_{\mu}}\left[
\int_{0}^{\infty}du_{i}p(u_{i})\right]\delta(u-\sum_{i=1}^{s_{\mu}}u_{i})
\nonumber \\
&=&\frac{1}{\Gamma(\mu)}
u^{\mu-1}e^{-u}.  
\end{eqnarray}
$U_{i}^{\mu}$ obeys a gamma distribution
with the shape exponent $s_{\mu}$  given by 
({\ref{Eq:Gamma}}). We note that the result (\ref{Eq:Gamma}) is 
derived in the thermodynamic limit, where the correlation among
$\{u_{j}\}_{j=1,\cdots k}$ vanishes. On the other hand, in the
continuous time branching process, the splitting processes of each individual
and offspring are independent of each other. 
As a result, we obtain the 
gamma distribution which appears in the voting model in 
the thermodynamic limit.

\end{document}